\documentclass[preprint,aps,amsmath,nofootinbib,tightenlines]{revtex4}
\usepackage{graphicx}
\usepackage{bm}
\usepackage{epsfig}
\usepackage{graphics}
\usepackage{color}

\newif\ifpdf
\ifx\pdfoutput\undefined
\pdffalse 
\else
\pdfoutput=1 
\pdftrue
\fi

\def\Dsl{\hbox{/\kern-.6000em D}} 

\def\dsl{\,\raise.15ex\hbox{/}\mkern-13.5mu D}

\def\ltap{\ \raise.3ex\hbox{$<$\kern-.75em\lower1ex\hbox{$\sim$}}\ }
\def\gtap{\ \raise.3ex\hbox{$>$\kern-.75em\lower1ex\hbox{$\sim$}}\ }
\def\OMIT#1{}

\def\lsim{\mathrel{\raise.3ex\hbox{$<$\kern-.75em\lower1ex\hbox{$\sim$}}}}
\def\gsim{\mathrel{\raise.3ex\hbox{$>$\kern-.75em\lower1ex\hbox{$\sim$}}}}

\def\tb#1{\textcolor{blue}{#1}}

\newcommand{\bmk}{\mathbf k}

\newcommand{\bbmp}{\mbox{\scriptsize\boldmath $p$}}

\catcode`\@=11
\def\slash{\mathpalette\make@slash}
\def\make@slash#1#2{\setbox\z@\hbox{$#1#2$}%
  \hbox to 0pt{\hss$#1/$\hss\kern-\wd0}\box0}
\catcode`\@=12 

\begin{document}
\ifpdf
\DeclareGraphicsExtensions{.pdf, .jpg}
\else
\DeclareGraphicsExtensions{.eps, .jpg}
\fi


\preprint{ \vbox{ \hbox{MPP-2006-33} 
}}

\title{\phantom{x}\vspace{0.5cm} 
On Electroweak Matching Conditions for Top Pair Production at Threshold
\vspace{1.0cm} }

\author{Andr\'e~H.~Hoang and Christoph~J.~Rei\ss er\vspace{0.5cm}}
\affiliation{Max-Planck-Institut f\"ur Physik\\
(Werner-Heisenberg-Institut), \\
F\"ohringer Ring 6,\\
80805 M\"unchen, Germany\vspace{1cm}
\footnote{Electronic address: ahoang@mppmu.mpg.de, reisser@mppmu.mpg.de}\vspace{1cm}}


\begin{abstract}
\vspace{0.5cm}
\setlength\baselineskip{18pt}
We determine the real parts of electroweak matching conditions relevant for top
quark pair production close to threshold in $e^+e^-$ annihilation at
next-to-next-to-leading logarithmic (NNLL) order. 
Numerically the corrections are comparable to the NNLL QCD corrections.
\end{abstract}
\maketitle


\newpage

%
%
%
\section{Introduction}
\label{sectionintroduction}
The detailed study of top quark pair production close to threshold constitutes
a major part of the top quark physics program at the International Linear
Collider (ILC) during its first phase running at lower energies.  While the
top quark mass (in a threshold mass
scheme~\cite{synopsis,Brambilla:2004wf,CKMworkshop}) 
can be measured with a  
precision of around 100~MeV from the c.m.\,energy where the cross section
line-shape rises, accurate measurements of the top quark total width, the
strong coupling and (in the case of a light Higgs) the top Yukawa coupling can
be gained from the normalization and the form of the
line-shape~\cite{TTbarsim}. Electroweak effects play a crucial role for
theoretical 
predictions of top pair threshold production. Apart from the $t\bar t$
production process, mediated either by virtual photon or Z boson exchange, one
also needs to account for initial state QED beam effects and the finite top
quark lifetime in the leading order approximation. On the other hand, due to
the fact that, close to threshold, the top quarks are produced with small
relative velocities, $v\ll 1$, singularities proportional to $(\alpha_s/v)^n$
and $(\alpha_s\ln v)^n$ arise from virtual and real gluon radiation in
${\cal O}(\alpha_s^n)$ QCD diagrams that need to be summed up to all orders in
$\alpha_s$. These singularities arise from ratios of the kinematical scales
$m_t$ (``hard''), $p_t\sim m_t v\sim m_t\alpha_s$ (``soft'') and $E_t\sim m_t
v^2\sim m_t\alpha_s^2$ (``ultrasoft'') that are governing the nonrelativistic
$t\bar t$ dynamics, where $p_t$ and $E_t$ are the top three-momentum and kinetic
energy, respectively. Since the Standard Model top quark width
$\Gamma_t\approx \Gamma_t(t\to b W)\approx 1.5$~GeV is numerically comparable
to the average top kinetic energy, it serves as an infrared cutoff and
protects the nonrelativistic $t\bar t$ dynamics from 
nonperturbative QCD effects.

In recent years most work in the literature on top pair threshold production
was focused on determining higher order QCD corrections using nonrelativistic
QCD (NRQCD).~\footnote{We use the term
NRQCD to refer to a generic low-energy effective theory which describes
nonrelativistic $t\bar t$ pairs and bound state effects and not for a theory 
valid only for scales $m_t > \mu > m_t v$. 
}
In the fixed-order approach, where the singularities $\propto (\alpha_s/v)^n$
are systematically summed, next-to-next-to-leading order (NNLO) corrections to
the total cross section are fully known~\cite{synopsis} and even some NNNLO
analyses have become available recently (see
e.g. Refs.~\cite{Penin:2005eu,Beneke:2005hg}). In the renormalization group  
improved approach both singularities $\propto (\alpha_s/v)^n$ and $\propto
(\alpha_s \ln v)^n$ are systematically summed. Here, the full set of NNLL
order  corrections is known except for the full NNLL running of the
dominant S-wave $t\bar t$ production current~\cite{hmst,hmst1,Hoang3loop}. At
present the theoretical QCD normalization
uncertainties of the NNLO fixed-order prediction for the total $t\bar t$ cross
section close to threshold is estimated to be of the order 20\%~\cite{synopsis},
while for the (yet incomplete) NNLL order renormalization group
improved prediction it is 6\%~\cite{HoangEpiphany}. 

As far as electroweak corrections are concerned no complete determination of
all effects beyond the leading order level has yet been achieved for the total
cross section. At leading order the three electroweak effects mentioned above
can be treated independently: The top pair production is described by 
$(e^+e^-)(t\bar t)$ effective NRQCD operators with Wilson coefficients
that account for the propagation of the virtual photon and the Z boson 
mediating
the $t\bar t$ production process. The QED beam effects consisting of the
machine-dependent beam energy spread, beamstrahlung and initial state
radiation can be conveniently implemented through a luminosity spectrum that
is convolved into the theoretical predictions without any initial state QED
corrections~\cite{TTbarsim,Boogert:2002jr,Cinabro1}. The top quark width, on
the other hand, is incorporated 
into NRQCD by an imaginary matching condition of a bilinear top quark mass
operator and can be effectively accounted for by shifting the c.m.\,energy
into the complex plane,
$\sqrt{s}\to\sqrt{s}+i\Gamma_t$~\cite{Kuehn1,Fadin1}. In
Ref.~\cite{HoangReisser1} a systematic procedure for the theoretical treatment
of the effects from the finite top quark lifetime for the total cross section
beyond leading logarithmic (LL) order in the renormalization group improved
approach was presented. It was shown that the effects of the top decay can be
systematically incorporated when NRQCD is matched to the Standard Model by
including the absorptive parts in electroweak matching conditions that are
related to the top quark decay. A systematic expansion scheme can be achieved
by using the power counting~\cite{HoangReisser1,HoangTeubnerdist}
\begin{equation}
\label{powercounting}
v\, \sim\,\alpha_s\,\sim\, g\,\sim\, g^\prime
\end{equation} 
for $\alpha_s$ and the SU(2) and U(1) gauge couplings $g$ and
$g^\prime$. The parametric  
counting scheme in Eq.~(\ref{powercounting}) is motivated by the close numeric
relation between the average top quark kinetic 
energy and the top quark width, $E_t\sim m_t\alpha_s^2\approx \Gamma_t\sim
m_t\alpha$. It shows that one-loop electroweak corrections to the $t\bar t$
production process can contribute at NNLL order. In particular, it was shown
in Ref.~\cite{HoangReisser1} that, 
due to gauge cancellations, there are no electroweak matching corrections to
the NRQCD potential operators and heavy-quark-gluon interactions up to NNLL
order. However, nontrivial (real and imaginary) electroweak matching
conditions exist for the dominant $(e^+e^-)(t\bar t)$ operators as well as for the top
bilinear operators due to higher order corrections to the top quark width
and the relativistic time dilatation effects. In Ref.~\cite{HoangReisser1} the
imaginary parts of these electroweak matching conditions were analyzed in
detail. It was found that they lead to corrections of the line-shape form
between 2\% and 10\% and can shift the 1S peak position by up to 50~MeV.

In this work we analyze the real parts of the NNLL electroweak matching
conditions of the dominant S-wave $(e^+e^-)(t\bar t)$ operators. We exclude
pure QED corrections\footnote{In this context pure QED corrections
  correspond to loop diagrams that contain a photon but no other Standard
  Model gauge bosons.}
 from our considerations. A complete treatment of pure QED
effects together with a consistent computation of NNLL initial state (beam)
and final state corrections shall be postponed to subsequent work. The
corrections considered in this work have been discussed in detail before in
Ref.~\cite{Grzadkowski:1986pm} and were partly reanalyzed again later in
Ref.~\cite{GuthKuehn}. Here we 
present the results of a new independent computation and point out a number of
discrepancies to the results in these references.

The program of this paper is as follows. In
Sec.\,\ref{sectionnotation} we explain the notation relevant for the matching
conditions of the $(e^+e^-)(t\bar t)$ operators considered in this work. In
Sec.\,\ref{sectionmatching} we present our results and discuss the differences
to Refs.~\cite{Grzadkowski:1986pm} and~\cite{GuthKuehn}. Section~\ref{sectionanalysis}
contains a brief 
numerical analysis, and in Sec.\,\ref{sectionconclusion} we conclude.

\section{Notation} 
\label{sectionnotation}

In this section we set up the notation for the electroweak matching
conditions to the dominant S-wave $t\bar t$ production and annihilation
operators relevant for $e^+e^-$ collisions. To be specific we use the notations that
were introduced in Ref.~\cite{HoangReisser1} in the framework of
vNRQCD~\cite{LMR,HoangStewartultra}. The 
dominant effective theory operators to be used for $t\bar t$ pair production in
an S-wave spin triplet state have the form 
\begin{eqnarray}
{\cal O}_{V,\bbmp} & = & 
\big[\,\bar e\,\gamma_j\,e\,\big]\,{\cal O}_{\bbmp,1}^j
\,,
\nonumber
\\[2mm]
{\cal O}_{A,\bbmp} & = &
\big[\,\bar e\,\gamma_j\,\gamma_5\,e\,\big]\,{\cal O}_{\bbmp,1}^j
\,,
\label{OVAdef}
\end{eqnarray}
where
\begin{eqnarray}
{\cal O}_{\bbmp,1}^j & = &
\Big[\,\psi_{\bbmp}^\dagger\, \sigma_j (i\sigma_2)
  \chi_{-\bbmp}^*\,\Big]
\,
\end{eqnarray}
and the indices $j=1,2,3$ are summed.
They give the contribution 
$\Delta {\cal L} = \sum_{\bbmp} \left(C_V(\nu)\,{\cal O}_{V,\bbmp} +  C_A(\nu)\,
{\cal O}_{A,\bbmp} \right) + \mbox{H.c.}$ to the effective theory Lagrangian
where the hermitian conjugation is referring to the operators only and does,
due to unitarity, not act on the complex Wilson
coefficients~\cite{HoangReisser1}. The 
term $\nu$ is the dimensionless vNRQCD renormalization group parameter, where
$\nu=1$ corresponds to the hard matching scale ($\mu=m_t$) and
$\nu\sim\alpha_s$ indicates the range of scales where the effective theory
matrix elements are evaluated. 
The relation between the Wilson coefficients at the matching scale $\nu=1$ and
for $\nu<1$ has the multiplicative form
$C_{V,A}(\nu)/C_{V,A}(1)=\exp(f(m_t,\nu))$. The function $f$ is fully known at
NLL order~\cite{HoangStewartultra,Pineda1}. At NNLL order only the nonmixing
contributions 
to the anomalous dimensions~\cite{Hoang3loop} and the subleading mixing
effects from the spin-dependent potentials~\cite{Penin:2004ay} are presently
known. The matching conditions for the 
Wilson coefficients at NNLL order (and without accounting for pure QED
corrections) can be written in the form ($i=V,A$)
\begin{eqnarray}
\label{matchingC}
C_i(\nu=1) & = & 
C_i^{\rm born}(\nu=1)\,\left(\,
1-\frac{2 C_F\alpha_s(m_t)}{\pi} + \Big(\frac{\alpha_s(m_t)}{\pi}\Big)^2 a^{(2)}
\,\right)
+ i\,C_i^{\rm bW, abs} + C_i^{\rm ew}\,.
\end{eqnarray} 
The terms $C_{V,A}^{\rm born}$ refer to the tree level matching
conditions originating from virtual photon and Z boson exchange and read
\begin{eqnarray}
C_V^{\rm born}(\nu=1) & = & 
\frac{\alpha\pi}{m_t^2(4c_w^2-x)}\,\bigg[\,
Q_e Q_t(4-x) + Q_t-Q_e-\frac{1}{4s_w^2}
\,\bigg]
\,,
\label{Cvborn}
\\[2mm]
C_A^{\rm born}(\nu=1) & = &
-\,\frac{\alpha\pi}{m_t^2(4c_w^2-x)}\,\bigg[\,
Q_t-\frac{1}{4s_w^2}
\,\bigg]
\,,
\label{Caborn}
\end{eqnarray}
where
\begin{equation}
x\equiv \frac{M_W^2}{m_t^2}\,,
\end{equation}
$Q_f$ is the fermion electric charge, $s_w$ ($c_w$) the sine (cosine) of
the weak mixing angle and $\alpha$ is the electromagnetic coupling. The factors
multiplying the tree level contributions represent the hard QCD matching
conditions which are presently known up to order $\alpha_s^2$.
The term $C_{V,A}^{\rm bW, abs}$ in Eq.~(\ref{matchingC}) describes the
imaginary (absorptive) 
parts of the matching conditions. They arise from cuts in the one-loop
electroweak diagrams that are related to intermediate states that arise in
the top decay and were identified in Ref.~\cite{HoangReisser1}. In predictions
for the total cross section they account for interference contributions of the
amplitude for the double resonant signal process $e^+e^-\to t\bar t\to 
bW^+\bar b W^-$ with the single
resonant amplitudes describing the processes $e^+e^-\to t+\bar b W^-\to
bW^+\bar b W^- $  and 
$e^+e^-\to bW^+\bar t\to bW^+\bar b W^-$. 
According to the power counting scheme in Eq.~(\ref{powercounting}) these
interference effects contribute at NNLL order to the NRQCD matrix elements for
the total cross section. In Ref.~\cite{HoangReisser1} it was also shown that these NNLL
matrix elements contain phase space UV-divergences that lead to the NLL
running of  $(e^+e^-)(e^+e^-)$ forward scattering operators. Finally, the terms
$C_{V,A}^{\rm ew}$ in Eq.~(\ref{matchingC}) refer to the real parts of the
NNLL (one-loop) electroweak matching corrections, which are the main focus of
this work.  

Using the optical theorem the contribution of the effective theory operators
in  Eqs.~(\ref{OVAdef}) to the total $t\bar t$ production cross section close
to threshold reads
\begin{eqnarray}
\sigma_{\rm tot} & \sim &
\frac{1}{s}\, 
\mbox{Im}\left[\,\Big(C_V^2(\nu)+C_A^2(\nu)\Big)\,L^{lk}\,{\cal  A}_1^{lk}\,\right]
\,,
\label{opttheo}
\end{eqnarray}
where ($k+k^\prime=(\sqrt{s},0)$ and ${\bf \hat e}=\bmk/|\bmk|$) 
\begin{eqnarray}
L^{lk} &=&
       \frac{1}{4}\,\sum\limits_{e^\pm\rm spins}\,
       \Big[\,\bar v_{e^+}(k^\prime)\,\gamma^l\,(\gamma_5)\,u_{e^-}(k)\,\Big]
       \,\Big[\,\bar u_{e^-}(k)\,\gamma^k\,(\gamma_5)\,v_{e^+}(k^\prime)\,\Big]
\nonumber
\\[2mm]
       &=& \frac{1}{2}\,(k+k^\prime)^2\,(\delta^{lk}-\hat e^l  \hat e^k)
\end{eqnarray}
is the spin- and angular-averaged lepton tensor and
($\hat{q}\equiv(\sqrt{s}-2m_t,0)$) 
\begin{eqnarray}
{\cal A}_1^{lk} &=& 
i\, \sum\limits_{\mbox{\scriptsize\boldmath $p$},\mbox{\scriptsize\boldmath $p'$}} \int\! d^4x\:
e^{-i\hat q \cdot x}\:
\Big\langle\, 0\,\Big|\,T\,
{{\cal O}_{\mbox{\scriptsize\boldmath $p$},1}^l}^{\!\!\!\dagger} (0)\, 
{\cal O}_{\mbox{\scriptsize\boldmath $p'$},1}^k (x)\Big|\,0\,\Big\rangle
\end{eqnarray}
is the time ordered product of the $t\bar t$ production and
annihilation operators ${\cal O}_{\bbmp,1}^j$ and ${{\cal
    O}_{\bbmp,1}^j}^{\!\!\!\dagger}$, which describes the
nonrelativistic dynamical effects. 
In the case of nonzero electron or positron polarization the formula in
Eq.~(\ref{opttheo}) is simply modified by 
\begin{eqnarray}
C_V^2+C_A^2\, \stackrel{P_{+},P_{-}\neq 0}{\longrightarrow}\,
(1-P_{+}P_{-})\,( C_V^2+C_A^2 ) + (P_{-}-P_{+})\,2\, C_V\, C_A
\,,
\end{eqnarray} 
where $P_{\mp}$ refers to electron/positron polarization along the momentum
direction. Since $C_V$ and $C_A$ have opposite signs (see Tab.~\ref{tabc})
the cross section is large for positive $e^+$ and negative $e^-$ polarization.
In the unpolarized case the relative NNLL corrections to the total cross
section coming from the real parts of the one-loop electroweak matching
conditions read
\begin{eqnarray}
\Delta^{\rm ew} 
\, = \, \frac{\delta \sigma_{\rm tot}^{\rm ew}}{\sigma_{\rm tot}}
\, = \, \frac{2C_V^{\rm born} C_V^{\rm ew}+2C_A^{\rm born} C_A^{\rm ew}}
{(C_V^{\rm born})^2+(C_A^{\rm born})^2}
\,,
\label{deltaew}
\end{eqnarray}
where all coefficients can be taken at $\nu=1$.

\section{Matching Procedure and Results} 
\label{sectionmatching}

The NNLL order electroweak matching conditions for the Wilson coefficients of
the effective theory operators in Eqs.~(\ref{OVAdef}) are obtained from the
one-loop electroweak corrections to the Standard Model amplitude for
$e^+e^-\to t\bar t$ for on-shell external top quarks in the limit $\sqrt{s}\to
2m_t$ (i.e. $v\to 0$), where the top quarks are at rest. 
For simplicity the electron mass is neglected except for the self
energy corrections, the CKM matrix is considered to be the unit matrix and
the bottom quark mass is neglected in the $\gamma t\bar t$ and $Zt\bar
t$ vertex corrections (including the top quark wave function renormalization).
If QED corrections are
neglected, there are no contributions singular in $v$ that are associated with
nontrivial NRQCD matrix elements and all contributions need to be absorbed
directly into the effective theory matching conditions. Neglecting
pure QED corrections the Standard Model amplitude can be written in the
form
\begin{eqnarray}
{\cal A} & = &
i\,\frac{\alpha\,\pi}{m_t^2}\,
\Big[\,\bar v_{e^+}(k^\prime)\,\gamma^\mu
 (h_R^{\rm ew}\,\omega_{+} + h_L^{\rm ew}\,\omega_{-})\, u_{e^-}(k)\,\Big]\,
\Big[\,\bar u_t(p)\,\gamma_\mu\,v_{\bar t}(p)\,\Big]
\,,
\label{eett}
\end{eqnarray}
where $\omega_{\pm}\equiv \frac{1}{2}(1\pm\gamma_5)$ and
$k+k^\prime=2p=(2m_t,0)$.
The real parts of the NNLL electroweak matching conditions then read
\begin{eqnarray}
C_V^{\rm ew}(\nu=1) \, = \, \frac{\alpha\,\pi}{2\,m_t^2}\,\mbox{Re}[\,h_R^{\rm ew}+h_L^{\rm ew}\,]
\,,
\nonumber\\[2mm]
C_A^{\rm ew}(\nu=1) \, = \, \frac{\alpha\,\pi}{2\,m_t^2}\,\mbox{Re}[\,h_R^{\rm ew}-h_L^{\rm ew}\,]
\,.
\label{Cewdef}
\end{eqnarray}
To our knowledge the coefficients $h_{L,R}^{\rm ew}$ were first computed in
Ref.~\cite{Grzadkowski:1986pm} in Feynman gauge using the on-shell scheme for the initial state
$e^+e^-$ pair, the final state $t\bar t$ pair and the intermediate Z boson,
and employing the fine structure constant for the electromagnetic coupling 
$\alpha^{-1}=137.036$. We have determined the real parts of the coefficients
$h_{L,R}$ in the same renormalization scheme by hand and, independently, using
the automated packages FeynArts~\cite{Hahn:2000kx} and
FormCalc~\cite{Hahn:1998yk} (except for the $Z\gamma$ box diagrams). For the 
presentation of our results we follow (with minimal modifications) the
conventions and notations used in Ref.~\cite{Grzadkowski:1986pm}. Since all
components of the coefficients $h_{L,R}$ were presented explicitly in
Ref.~\cite{Grzadkowski:1986pm} we discuss in detail only those
elements where our results differ. 
For undefined symbols and variables used in
the following discussion we refer the reader to Ref.~\cite{Grzadkowski:1986pm}.

The results for the coefficients $h_{L,R}^{\rm ew}$ can be cast into the form
\begin{equation}
h_{L,R}^{\rm ew}\, = \,
h_{L,R}^{\rm SE}\,+\, h_{L,R}^{e^+e^-}\,+\,
h_{L,R}^{t\bar t}\,+\,h_{L,R}^{\rm box}
\,,
\end{equation}
where ($M^2=s=4m_t^2$)
\begin{eqnarray}
h_{L,R}^{\rm SE} & = & Q_e \Big(-\frac{\Pi_R^{AA}}{M^2}\Big) Q_t 
        + \beta_{L,R}^e \frac{M^2}{M^2-M_Z^2}
        \Big(-\frac{\Pi_R^{ZZ}}{M^2-M_Z^2}\Big)
        \frac{\beta_R^t+\beta_L^t}{2}
\nonumber\\
    &&{}- \Big(Q_e \frac{\beta_R^t+\beta_L^t}{2} + \beta_{L,R}^e Q_t\Big)
        \frac{\Pi_R^{ZA}}{M^2-M_Z^2}
\,,
\\[2mm] 
h_{L,R}^{e^+e^-} & = & \tb{F_{L,R}^A} Q_t + \tb{F_{L,R}^Z} \frac{M^2}{M^2-M_Z^2}
        \frac{\beta_R^t+\beta_L^t}{2}
\,,
\\[2mm] 
h_{L,R}^{t\bar t} & = & Q_e \frac{\alpha}{4\pi} \tb{\sum a^A} + \beta_{L,R}^e
	\frac{M^2}{M^2-M_Z^2} \frac{\alpha}{4\pi} \tb{\sum a^Z}
\,,
\\[2mm] 
h_{L,R}^{\rm box} & = & h_{L,R}^{WW} + h_{L,R}^{ZZ} + \tb{h_{L,R}^{Z\gamma}}
\,
\end{eqnarray}
and
\begin{eqnarray}\label{suma}
\sum a^{A,Z} &\equiv& a^{A,Z}(W) + a^{A,Z}(W,W) + a^{A,Z}(\phi_W) +
          a^{A,Z}(\phi_W,\phi_W)
\nonumber\\[2mm]
	  &&{}+ a^{A,Z}(W,\phi_W) + a^{A,Z}(Z) + a^{A,Z}(\phi_Z) +
          \tb{a^{A,Z}(H)}
\nonumber\\[2mm]
	  &&{}+ a^{A,Z}(Z,H) +
          a^{A,Z}(\phi_Z,H)+\tb{a^{A,Z}_{\rm ct}}
\,
\end{eqnarray}
and
\begin{equation}
\beta_R^f \, = \, -\frac{s_w^2\,Q_f}{s_w c_w}
\,,\quad
\beta_L^f \, = \, \frac{t_{3,f}-s_w^2 Q_f}{s_w c_w}
\,, 
\end{equation}
$t_{3,f}$ being the third component of the weak isospin of fermion $f$.

The terms $h_{L,R}^{\rm SE}$ describe the self energy corrections where
$\Pi_R^i$, $i=AA,ZZ,ZA$, are the renormalized transverse photon and Z
self energy corrections and the photon-Z mixing correction. Compared to the
expressions given in Ref.~\cite{Grzadkowski:1986pm} our result does not sum the photon and Z
boson self energy corrections into the denominator of the respective
propagators since the resulting higher order corrections are beyond NNLL order
and because at the $t\bar t$ threshold the intermediate photon and Z boson are
far off-shell. 
The terms  $h_{L,R}^{e^+e^-}$ describe the corrections to
the $e^+e^-$ vertex, where $F_{L,R}^i$, $i=A,Z$, are the vertex corrections to
the left- and right-handed $e^+e^-\gamma$ and  $e^+e^-Z$ vertices. 
The terms  $h_{L,R}^{t\bar t}$ refer to the corrections to the $t\bar t$
vertices, where the coefficients $a^{A,Z}$ correspond to the corrections to
the $\gamma t\bar t$ and $Z t\bar t$ vertices, respectively. The arguments of
the various terms in Eq.~(\ref{suma}) indicate the virtual bosons that are being
exchanged in the various triangle diagrams, e.g. $a^{A,Z}(W,W)$ refers to the
vertex diagrams with exactly two internal W lines and $a^{A,Z}(\phi_Z)$ to those where the
only internal boson is the neutral pseudo-Goldstone boson. Note that
$a^{A,Z}(W,\phi_W)$, $a^{A,Z}(\phi_Z,H)$ and $a^{A}(Z,H)$ are identically zero
in Feynman gauge. 
Finally, the terms $h_{L,R}^{\rm box}$ describe the contributions from the
$WW$, $ZZ$ and $Z\gamma$ box diagrams. We note that the $Z\gamma$ box diagrams
are infrared-finite for the $t\bar t$ pair being at rest. 

Our results agree with those given in Ref.~\cite{Grzadkowski:1986pm} except for
the following points:
\footnote{
We acknowledge that the discrepancies 2) and 3) were
confirmed to us by~\cite{privcom}. 
}
\begin{enumerate}
\item 
For the corrections to the left-handed $e^+e^-\gamma$ and $e^+e^-Z$ vertices
the coefficients $F_L^{A,Z}$ read
\begin{eqnarray}
F_L^A & = & \frac{\alpha}{4\pi}\left( Q_e \left(\beta_L^e\right)^2
  \rho(q,M_Z)- \frac{t_3^e}{s_w^2} \tb{\Lambda(q,M_W)}\right)
\,,
\\[2mm]
F_L^Z & = & \frac{\alpha}{4\pi}\left(\left(\beta_L^e\right)^3 \rho(q,M_Z)
  + \Big(\beta_L^e - 2 t_3^e \frac{c_w}{s_w}\Big)\frac{1}{2s_w^2} \rho(q,M_W)
  - \frac{c_w}{s_w}\frac{t_3^e}{s_w^2}
  \tb{\Lambda(q,M_W)}\right)
.
\end{eqnarray}
We agree with the result for the function $\rho$ in~\cite{Grzadkowski:1986pm}, but
for the function $\Lambda$ we find
\begin{eqnarray}
\Lambda(q,M_W) & = & -\frac52 +
\frac{2}{u}\,\tb{-}\left(1+\frac{2}{u}\right)l
\sqrt{1-\frac{4}{u}}\,\tb{-}\left(1+\frac{1}{2u}\right)\frac{4}{u}l^2
\,,
\end{eqnarray}
where
\begin{eqnarray}
l\equiv \ln\frac{\sqrt{1-4/u}+1+i \epsilon}{\sqrt{1-4/u}-1+i \epsilon}
=\ln \frac{1+\sqrt{1-4/u}}{1-\sqrt{1-4/u}}-i\pi\,,\quad
u\equiv\frac{M^2}{M_W^2}
\,.
\end{eqnarray}
For 
$\Delta^{\rm ew}$, which describes the corrections to the total cross
section from the real parts of the NNLL order electroweak matching conditions,
this leads to an absolute shift by $+0.012$ with respect to the results in
Ref.~\cite{Grzadkowski:1986pm}. 
\item 
For the corrections to the $t\bar t$ vertices coming from the exchange of the
neutral physical Higgs boson we find the result
\begin{eqnarray}
a(H) & = & \binom{-Q_t}{-\tfrac12(\beta_R^t + \beta_L^t)} 
           \frac{1}{s_w^2} \frac{M^2}{48M_W^2} \Bigg\{1 +
           \left(3\frac{M^2}{M_H^2}-1\right) B_0(\tfrac12 P, M_H, \tfrac12 M)
\nonumber\\[2mm]
        &&{}+ B_1(\tfrac12 P,M_H,\tfrac12 M)-3\frac{M^2}{M_H^2}
           \tb{B_0(P,\tfrac12 M,\tfrac12 M)}\Bigg\}
\,,
\end{eqnarray} 
where the upper component refers to the photon and the lower to the Z vertex,
respectively. Our results are consistent with the ones in
Refs.~\cite{GuthKuehn,Jezabek:1993tj}, but differ from those in Ref.~\cite{Grzadkowski:1986pm} in
the argument of the last $B_0$ function. Furthermore the
counterterm contributions read
\begin{eqnarray}
a^A_{\rm ct} & = & \frac{2\pi}{\alpha}(Q_t(\tb{\delta Z_R}+\tb{\delta Z_L})+2 t_3^t\,\delta k)
\,,
\\[2mm]
a^Z_{\rm ct} & = & \frac{2\pi}{\alpha}\left((\beta^t_R
\tb{\delta Z_R}+\beta^t_L \tb{\delta Z_L})+2
\frac{c_w}{s_w} t_3^t\,\delta k \right)
\,.
\end{eqnarray}
For the left- and right-handed wave function renormalization terms we have the
expression 
\begin{eqnarray}
\delta Z_{L,R} & = & A_{L,R} + m_t^2 (A_R^\prime + A_L^\prime - 2\tb{C^\prime})
\,,
\end{eqnarray}
where $C^\prime \equiv \frac{dC}{dp^2}(p^2=m_t^2)$. For the function $C$ we find
\begin{eqnarray}
C & = & \frac{\alpha}{4\pi
  s_w^2} \Big\{2s_w^2\beta^t_R\beta^t_L(1 - 2\tb{B_0(p,m_t,M_Z)})
  +\frac{m_t^2}{4M_W^2}\big(B_0(p,m_t,M_H) 
\nonumber\\
   &&{}- B_0(p,m_t,M_Z)\big)\Big\}
\,,
\end{eqnarray}
which is consistent with Ref.~\cite{GuthKuehn}, but differs from Ref.~\cite{Grzadkowski:1986pm}, where the $B_1$ function
was obtained instead of the first $B_0$ function. For 
$\Delta^{\rm ew}$ 
these
changes lead to an absolute shift by $+0.076$ with respect to the results in
Ref.~\cite{Grzadkowski:1986pm} for $m_H=130$~GeV. 
\item 
For the corrections from the $Z\gamma$ box diagrams we find the result
\begin{eqnarray}
h_R^{Z\gamma} & = & \frac{\alpha}{4\pi}(-\beta_R^e)(\beta_R^t-\beta_L^t)Q_eQ_t
2\tb{F^{Z\gamma}}
\,,
\\[2mm]
h_L^{Z\gamma} & = & \frac{\alpha}{4\pi}(\beta_L^e)(\beta_R^t-\beta_L^t)Q_eQ_t
2\tb{F^{Z\gamma}}
\,,
\end{eqnarray}
where ($\tilde u\equiv\frac{M_Z^2}{M^2}$)
\begin{eqnarray}
F^{Z\gamma} & = & \tb{-}\frac{1}{\tilde u - 1} (B_0(\tfrac12 P,\tfrac12
  M,0) + B_0(\tfrac12 P,\tfrac12 M,M_Z) - 2 B_0(P,0,M_Z))
\nonumber\\
  &&{}+ (B_0(P,0,M_Z) - B_0(l,0,\tfrac12 M)) + (\tilde u - 1) M^2
  C_0(l,\tfrac12 P,0,\tfrac12 M,M_Z)
\,.
\end{eqnarray}
The expression for $F^{Z\gamma}$ agrees with the one in
Ref.~\cite{Stuart:1987tt}, but differs from the one in
Ref.~\cite{Grzadkowski:1986pm} by an 
additional minus sign in front of the first term. We note that the same error
is also contained in the analysis of Ref.~\cite{GuthKuehn}.  For 
$\Delta^{\rm ew}$ 
the
changes lead to an absolute shift by $+0.004$ with respect to the results in
Ref.~\cite{Grzadkowski:1986pm}.   
\end{enumerate}

At this point we also take the opportunity to point out a number of
discrepancies we find to Ref.~\cite{GuthKuehn}. (For the definition of the
various undefined variables we refer the reader to Ref.~\cite{GuthKuehn}.) For
the WW box contribution we find
\begin{eqnarray}
F^{WW}(r) & = & \frac{2}{r - 1}[l_1 - l_3 + \tfrac14 (f_2 -
  \tb{r f_1})]
\,,
\end{eqnarray}
where the function $f_1$ is defined by
\begin{eqnarray}
f_1(r) & = &
\tb{-}\left(\ln\frac{1+\beta}{1-\beta}-i\pi\right)^2
\,, \quad \beta \equiv \sqrt{1-r}
\,.
\end{eqnarray}
This differs from~\cite{GuthKuehn} (in the published version) by the factor $r$
multiplying the function $f_1$ and the global minus sign in the
definition of $f_1$. For the contribution of the $Z\gamma$ box diagrams we
find the expression
\begin{eqnarray}
F^{Z\gamma}(r) & = & \tb{+} 2 \ln r + \ln
        \frac{4}{r} - \left(\tb{-1} - \frac{r}{4}\right)
        \left(\ln \left(\frac{4}{r} - 1\right) - i \pi\right)
        \tb{-}\,\frac{4}{4-r}\, l_4 + \left(\frac{r}{4} - 1\right)f_3
\,.
\end{eqnarray}
For the $t\bar t$ vertex correction $a^R_{ZH}$, which corresponds to
the triangle diagram containing a Z boson and a physical neutral Higgs
boson, we
find the expression
\begin{eqnarray}
a^R_{ZH}(r_H,r_Z) & = & \binom{0}{1} \frac{\beta_L^t + \beta_R^t}{4
  c_w^2} \Bigg\{
  \frac{r_H+r_Z}{8}\ln\frac{r_H}{r_Z}+\frac{1}{r_Z+r_H-4} 
  \Big[\tb{-}(4-r_H+r_Z)l_5 + (4-r_H)
\nonumber\\[2mm]
  &&{}\times \Big[\Big(1-\frac{r_H}{2}\Big)\tfrac12 \ln r_H r_Z -l_4^H\Big]
  +r_Z\Big[\Big(1-\frac{r_Z}{2}\Big)\tfrac12 \ln r_H r_Z - l_4^Z\Big]\Big]\Bigg\}
\,,
\end{eqnarray}
which differs from the one in~\cite{GuthKuehn} by a minus sign in front of the
function $l_5$.

\section{Numerical Analysis} 
\label{sectionanalysis}

\begin{table}
\begin{center}
\begin{tabular}{|l||r|r|r|r|r|r|r|r|}
\hline
$m_t$~(GeV)                          & \multicolumn{4}{|c|}{$170$} & \multicolumn{4}{|c|}{$175$}\\\hline
$m_H$~(GeV)                          & \multicolumn{1}{|c|}{$115$} & \multicolumn{1}{|c|}{$150$} & \multicolumn{1}{|c|}{$200$} & \multicolumn{1}{|c|}{$1000$} & \multicolumn{1}{|c|}{$115$} & \multicolumn{1}{|c|}{$150$} & \multicolumn{1}{|c|}{$200$} & \multicolumn{1}{|c|}{$1000$} \\\hline\hline
$C_V^{\rm born} (10^{-7}\,\mbox{GeV}^{-2})$& $-5.429$  &       & & & $-5.123$ & & & \\\hline
$C_A^{\rm born} (10^{-7}\,\mbox{GeV}^{-2})$& $1.260$   &       & & & $1.184$ & & & \\\hline
$C_V^{\rm ew} (10^{-8}\,\mbox{GeV}^{-2})$& $-4.562$ & $-4.073$ & $-3.702$ & $-3.060$ & $-4.460$ & $-3.951$ & $-3.566$ & $-2.890$\\\hline
$C_A^{\rm ew} (10^{-9}\,\mbox{GeV}^{-2})$& $-1.416$  & $-2.286$  & $-2.797$ & $-3.025$ & $-1.408$ & $-2.335$ & $-2.904$ & $-3.260$\\\hline
$C_V^{\rm ew}/C_V^{\rm born}$            & $0.0840$  & $0.0750$  & $0.0682$& $0.0564$ & $0.0871$ & $0.0771$ & $0.0696$ & $0.0564$\\\hline
$C_A^{\rm ew}/C_A^{\rm born}$            & $-0.0112$ & $-0.0181$ & $-0.0222$ & $-0.0240$ & $-0.0119$ & $-0.0197$ & $-0.0245$ & $-0.0275$\\\hline
\hline
\end{tabular}
\end{center}
{\caption{Numerical values for the tree level and real
    one-loop electroweak matching conditions $C_{V,A}^{\rm born}$ and
    $C_{V,A}^{\rm ew}$, respectively, and the $C_{V,A}^{\rm ew}/C_{V,A}^{\rm born}$ ratios for various values for the top
    and the Higgs masses, $\alpha^{-1}=137.036$ and the values given
    in Eq.~(\ref{values}). The coefficients $C_{V,A}^{\rm born}$ do not depend on the Higgs mass. }
\label{tabc} }
\end{table}

In this section we give a brief numerical discussion of the real parts of the
NNLL electroweak matching conditions obtained in this work. In Tab.~\ref{tabc}
the numerical values for  
$C_{V,A}^{\rm born}$ and $C_{V,A}^{\rm ew}$ are displayed for various values
for the top and the Higgs masses, $\alpha^{-1}=137.036$ and
\begin{eqnarray}
  \begin{array}{lll}
M_W=80.425~\mbox{GeV}\,, & M_Z=91.1876~\mbox{GeV}\,, & c_w^2=M_W^2/M_Z^2\,, \\
m_e=0.511~\mbox{MeV}\,,  & m_\mu=0.106~\mbox{GeV}\,, & m_\tau=1.78~\mbox{GeV}\,,\\
m_u=0.005~\mbox{GeV}\,, & m_d=0.005~\mbox{GeV}\,, & m_s=0.10~\mbox{GeV}\,,\\
m_c=1.3~\mbox{GeV}\,, & m_b=4.2~\mbox{GeV}\,
  \end{array}
\label{values}
\end{eqnarray}
for the gauge boson, the
lepton and the quark masses. Note that the finite electron mass is applied in the
calculation only for the self energy corrections.
The vector coefficients dominate for the tree level as well as for the
one-loop coefficients. The one-loop corrections show a significant Higgs mass
dependence and vary in the vector (axial-vector) case between 
$8.6\%$ ($-1.2\%$) and $5.6\%$ ($-2.6\%$) for Higgs masses between $115$~GeV
and $1$~TeV and $m_t=172.5$~GeV.

\begin{figure}[t] 
\begin{center}
\includegraphics[width=0.7\textwidth]{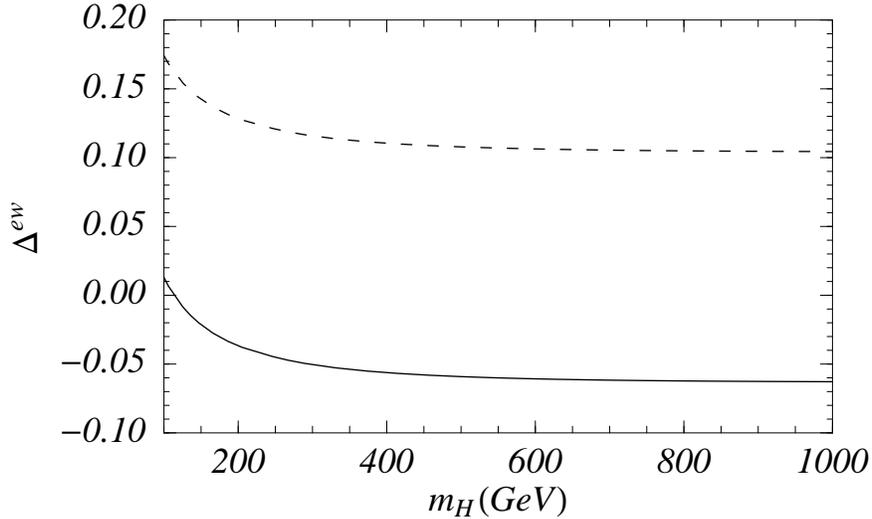}
 \caption{The correction $\Delta^{\rm ew}$ as a function of $m_H$ for
 $m_t=172.5$~GeV and the values given in Eq.~(\ref{values}) for
 $\alpha^{-1}=137.036$ (dashed curve) and in the
 scheme for the electromagnetic coupling defined in
 Eq.~(\ref{alphamsbar}) (solid curve).
 \label{figanalysis} }
\end{center}
\end{figure}

A more transparent view on the impact of the real electroweak one-loop
corrections on the predictions of the total $t\bar t$ threshold cross section
can be gained by considering the quantity $\Delta^{\rm ew}$ defined in
Eq.~(\ref{deltaew}). In Fig.~\ref{figanalysis} the dashed line
represents $\Delta^{\rm ew}$ as a
function of the Higgs mass for $m_t=172.5$~GeV and adopting the previous choices 
for the other parameters. 
%
%
%
For $m_H=(115,150,200,1000)$~GeV we have 
$\Delta^{\rm ew}=(0.161,0.142,0.128,0.104)$. The Higgs mass 
dependence is rather strong for small $m_H$ and drops quickly close to the
decoupling limit for increasing $m_H$. The strong Higgs mass dependence for
small Higgs masses originates from the fact that for a light Higgs boson the
dominant effects of the virtual Higgs exchange between the $t\bar t$ pair could
be described in NRQCD by a Yukawa potential that leads to a singularity
$\propto m_t/m_H$ for $m_H\to 0$ (see e.g. Refs.~\cite{GuthKuehn,Jezabek:1993tj,Harlander:1995dp}). In
an approach where the Higgs exchange would be accounted for in NRQCD through a
Yukawa potential the electroweak one-loop 
matching conditions would need to be modified  to account for the
contributions caused by the Yukawa potential in the NRQCD matrix elements. If
the Higgs boson is indeed very close to the lower LEP bound this would be a
viable alternative to our approach where all virtual electroweak effects are
encoded in the NRQCD Wilson coefficients. 

The solid line in Fig.~\ref{figanalysis}, finally, shows $\Delta^{\rm ew}$ for the
same choice of parameters but adopting an $\overline{\mbox{MS}}$ definition
for the QED coupling at the scale $\mu=m_t$ that accounts for the
leading-logarithmic vacuum polarization effects due to the three charged
leptons and the quarks below the top quark scale,
\begin{eqnarray}
\alpha^{n_f=8}(\mu)\, = \, 
\frac{\alpha}
{1-\frac{\alpha}{3\pi}\sum\limits_{i=e,\mu,\tau}Q_i^2
  \ln\left(\frac{\mu^2}{m_i^2}\right)
-\frac{\alpha}{3\pi}\sum\limits_{i=u,d,c,s,b}N_c Q_i^2
\ln\left(\frac{\mu^2}{m_i^2}\right) 
}
\,.
\label{alphamsbar}
\end{eqnarray}
In this scheme $\alpha^{n_f=8}(\mu=m_t)=1/125.926$ for $m_t=172.5$~GeV and the real
NNLL electroweak matching conditions are modified,
\begin{eqnarray}
C_{V,A}^{\rm ew,\overline{MS}} = C_{V,A}^{\rm ew} - C_{V,A}^{\rm
  born} \,
\frac{\alpha^{n_f=8}(\mu)}{3\pi}\left(
\sum\limits_{i=e,\mu,\tau}Q_i^2 \ln\left(\frac{\mu^2}{m_i^2}\right)
+\sum\limits_{i=u,d,c,s,b}N_c Q_i^2 \ln\left(\frac{\mu^2}{m_i^2}\right) 
\right)
\,.
\label{Cmod}
\end{eqnarray}
This scheme is preferred to the one where the fine structure
constant is used, since it leads to a substantially reduced dependence
on the light fermion masses in $\Delta^{\rm ew}$. This is because most
of the fermionic vacuum polarization effects are cancelled by the
modification in Eq.~(\ref{Cmod}). Moreover in this scheme 
$\Delta^{\rm ew}$ is by 17\% smaller than in the one with the fine
structure constant. For $m_H=(115,150,200,1000)$~GeV we now have  
$\Delta^{\rm ew}=(-0.001,-0.021,-0.037,-0.063)$.
The results show that concerning the real parts of the NNLL
order electroweak matching corrections (up to the QED effects
neglected in this work) the effects from virtual Higgs exchange
are comparable to the other electroweak effects.
\section{Conclusion} 
\label{sectionconclusion}

We have determined the real parts of the NNLL order (one-loop) electroweak
matching conditions for the Wilson coefficients of the dominant NRQCD
operators describing $t\bar t$ production close to threshold in $e^+e^-$
annihilation. The results include the contributions of all electroweak
one-loop effects that are integrated out when NRQCD is matched to the Standard
Model except for pure QED corrections. The results are an important ingredient
for a complete treatment of electroweak effects at NNLL order for top
quark pair production at a future Linear Collider. We have pointed out
a number of discrepancies with respect to earlier work.

\begin{acknowledgments} 
We would like to thank T.~Teubner and J.~H.~K\"uhn for comments to the manuscript.
\end{acknowledgments}

\newpage
\appendix


\end{document}